# Interpretation of Inelastic Neutron Scattering Data Using the Phase Diagram of Hole-Doped Cuprates

A. Mourachkine

*Université Libre de Bruxelles, Service de Physique des Solides, CP233, Boulevard du Triomphe, B-1050 Brussels, Belgium*

**Abstract**

Inelastic Neutron Scattering (INS) data in $La_{2-x}Sr_xCuO_4$ (LSCO), $YBa_2Cu_3O_{6+x}$ (YBCO) and $Bi_2Sr_2CaCu_2O_{8+x}$ (Bi2212) are discussed. In the literature, the INS spectra remain far from being comprehensively understood. We show that local ($Q$-integrated) susceptibility data and the energy dependence of the spin susceptibility at antiferromagnetic vector, $Q_{AF} = (\pi,\pi)$, can be interpreted by using the phase diagram for hole-doped cuprates. We analyze also the origin of the resonant peak which relates to the order parameter for long-range phase coherence in hole-doped cuprates. Thus, we present an interpretation of neutron data in LSCO, YBCO and Bi2212 published so far.

## 1. Introduction

Recent inelastic neutron scattering (INS) experiments have provided a considerable insight in the understanding of the anomalous properties of high-$T_c$ superconductors [1]. INS measurements provide essential information on the energy and momentum dependencies of the spin-spin correlation function [2-27]. The spin dynamics at low frequencies in SC cuprates are beginning to achieve the same universality across widely different materials with different values of $T_c$ [6,23]. The latter suggests that the magnetic fluctuations do not depend on the details of Fermi surface [20]. As for the bulk characteristics, the controlling factor in cuprates seems only to be the hole density in $CuO_2$ planes [6,23]. However, the INS spectra remain far from being comprehensively understood.

So-called magnetic "resonance peak" found earlier in $YBa_2Cu_3O_{6+x}$ (YBCO) [4] by INS has been recently detected in $Bi_2Sr_2CaCu_2O_{8+x}$ (Bi2212) [21]. The discovery of the resonance peak in Bi2212 points out that the resonance peak is an *intrinsic* feature of the superconductivity (SC) in the double-layer hole-doped cuprates studied so far. The resonance-peak position, $E_r$, increases first when the doping increases and then decreases again in the overdoped regime, thus, $E_r$ scales with $T_c$ [2,7,12,17,23]. The temperature dependence of the intensity of the resonance peak is very similar to the temperature dependence of a SC gap [7]. There is a consensus that the resonance peak is a consequence of the establishment of SC [2,3,10,12]. It is important to note that the magnetic energy stored in the resonance peak is in quantitative agreement with the SC condensation energy [28,29]. The resonance peak has been also observed by INS in a heavy fermion compound $UPd_2Al_3$ [30] for which spin fluctuations are believed to mediate the pairing interactions [31,32]. The SC in $UPd_2Al_3$ coexists with the antiferromagnetic (AF) order like in the cuprates [30-32]. By making a parallel between the cuprates and heavy fermion compound $UPd_2Al_3$, the latter suggests that spin fluctuations mediate the electron pairing that leads to SC in hole-doped cuprates.

The sharpness of the resonance peak close to optimal doping indicates a true collective mode [3]. There are many possible explanations for the origin of the magnetic resonance peak observed in hole-doped cuprates. On earlier stage, the resonance peak was associated with spin-flip charge carriers excitations across the SC gap [33]. Another explanation is that the magnetic resonance peak is due to electron-hole pair excitation across the SC energy gap [10]. It can be a collective spin-wave mode brought about by strong AF correlations [34]. More recently, the resonance peak was associated with the response of the magnetic AF domains to the hole pairing along charge stripes [35]. By analyzing tunneling and INS data [36] we proposed that the resonance peak corresponds to spin-waves excited in AF domains, which mediate the long-range phase coherence of the spinon SC which occurs along charge stripes (a Magnetic Coupling between Stripes (MCS) model) [37-39]. There are other explanations for the origin of the magnetic resonance peak in hole-doped cuprates [12,33].

It is important to note that an INS measurement is the signal averaging experiment. Regnault *et al.* first pointed out that INS spectra obtained at AF vector, $Q_{AF} = (\pi,\pi)$, are naturally characterized in the SC state by two distinct components [5]. The resonance peak becomes the major part of the spectrum with increasing doping. The two distinct contributions can be observed in both the energy and momentum dependencies of the spin susceptibility. However, the interpretation of INS spectra in terms of two contributions is not widely accepted [10]. Very recently, Fong *et al.* underlined that *Q-integrated* spin-susceptibility data consist also of two contributions [3]. The first component (resonant component) is exclusively observed in the odd channel of the spin susceptibility

(acoustic spin-waves) while the second component (spin gap or pseudogap) appears in the even channel (optical spin-waves). It seems that the spin gap can be observed at low temperatures also in the odd channel [8].

The purpose of this paper is to discuss the $Q$-integrated INS data and data obtained at AF vector, $Q_{AF} = (\pi,\pi)$, in $La_{2-x}Sr_xCuO_4$ (LSCO), YBCO and Bi2212. We show that by using the phase diagram of hole-doped cuprates it is possible to interpret the spin-susceptibility data. We discuss also the origin of the magnetic resonance peak.

## 2. The phase diagram of hole-doped cuprates

Before analyzing the INS spectra let us discuss first the phase diagram of hole-doped cuprates.

In conventional SCs, the SC order parameter (OP) is responsible simultaneously for pairing of two electrons (for the formation of the Cooper pairs) and for establishing the long-range phase coherence among these Cooper pairs. By other words, the pairing mechanism and mechanism of establishment of the long-range phase coherence in the classical SCs are identical: two electrons in each Cooper pair are attracted by phonons, and the phase coherence among the Cooper pairs is established also by phonons [40]. Both the phenomena occur almost simultaneously at $T_c$. In SC copper-oxides, there is a consensus that, at least, in the underdoped regime, the pairing occurs above $T_c$, when the long-range phase coherence is established, $T_{pair} > T_c$ [41-45]. Thus, this leads to the appearance of two distinct OPs in hole-doped cuprates, which are both SC-like [38]. One OP is for pairing and the second OP is for establishing the long-range phase coherence. The major reasons for the co-existence of the two OPs in copper oxides are the small value of coherence length (~ 15 - 20 Å) and low density of the Cooper pairs [44]. In conventional superconductors, the coherence length is usually very large (~ 500 - 16000 Å), and the density of the Cooper pairs is much higher than that in the cuprates [44].

The two OPs have been experimentally observed in tunneling measurements on pure and Ni-doped Bi2212 [36]. Independently from these measurements, the analysis of Andreev reflection, penetration depth, Raman scattering, angle-resolved photo-emission and tunneling measurements made by Deutscher showed the presence of two distinct energy scales in hole-doped cuprates [45]. Surprisingly, the two phase diagram obtained independently from each other practically coincide. The magnitudes of the OPs for pairing and long-range phase coherence in hole-doped cuprates have different dependencies on hole concentration in CuO$_2$ planes, $p$. Fig. 1 shows the phase diagram of the two OPs, which is *common* for hole-doped cuprates [36,45,46]. In Fig. 1, the magnitude of the OP for long-range phase coherence, $\Delta_c$, scales with $T_c$ as $2\Delta_c/k_B T_c = 5.45$ [45]. The magnitude of the OP for pairing, $\Delta_p$, increases linearly with the decrease of hole concentration. In the present work, we do not discuss the phases of the two OPs, which are locked to each other below $T_c$.

The consequence of the presence of pairing and coherent SC OPs is peculiar. In conventional SCs, these two OPs simply coincide. It is not easy to visualize the picture with the presence of pairing and coherent OPs. The presence of the two OPs implies that the total condensation energy of a Cooper pair in the SC state is equal to $E_{cond} = (\Delta_c^2 + \Delta_p^2)^{1/2}$ [47]. The latter depends not only on the symmetries of the two OPs but also on the relative angle between the OPs. If both $\Delta_c$ and $\Delta_p$ OPs have the isotropic s-wave symmetry, then in order to break a Cooper pair it is necessary to spend the energy $E_{cond}$ which does not depend on the angle. If one of the two OPs has the $d_{x^2-y^2}$ symmetry, which is the case in the cuprates [46,48], in order to break a Cooper pair it is necessary to spend, in some directions on the Fermi surface, the energy equal to $E_{cond}$, in some directions, only the energy equal to the magnitude of the s-wave OP [46]. If both OPs have the $d_{x^2-y^2}$ symmetry, then in order to break a Cooper pair the energy between 0 and $E_{cond}$ is needed, depending on the angle on the Fermi surface. In our previous studies, we considered in detail the symmetries of the two OPs. We found that, most likely, the OP for long-range phase coherence has the d-wave symmetry while the OP for pairing has an anisotropic s-wave symmetry [46,38].

We successfully used the phase diagram shown in Fig. 1 to interpret tunneling measurements. There is a clear discrepancy among the energy-gap values for different 90 K cuprates, inferred from tunneling measurements [49]. In Bi2212, there is a distribution of the gap magnitude (21–37 meV). At the same time, the maximum magnitudes of tunneling gaps in Tl$_2$Ba$_2$CuO$_{6+x}$ (Tl2201) (22 meV) and YBCO (21 meV) correspond to the minimum gap magnitude in Bi2212. However, all three cuprates have similar values of $T_c$. By using the phase diagram of hole-doped cuprates we showed that tunneling measurements performed on 90 K cuprates detect two different gaps (OPs) [49].

Let us apply the phase diagram shown in Fig. 1 for the interpretation of magnetic spectra obtained by INS on hole-doped cuprates.

## 3. Spin susceptibility at (π, π)

In this Section, we discuss the energy dependence of the spin susceptibility at AF wave-vector, $Q_{AF} = (\pi,\pi)$, in YBCO and Bi2212. The resonance peak appears in the odd channel of the spin susceptibility at $Q_{AF}$. At each doping, the peak intensity at the resonance energy is characterized by a striking temperature dependence displaying a pronounced kink at $T_c$ [10]. With increasing doping, the resonance peak becomes the major part of the magnetic spectrum. Besides the resonance peak in the INS spectra at (π,π), there is a non-resonant contribution which has been not received much attention so far [10]. With increasing doping, the latter peak is continuously reduced: it becomes too weak to be measured in INS experiments in the overdoped regime YBCO$_7$. In fact, in overdoped regime, all INS spectra become very weak. Above $T_c$, the spin susceptibility at (π,π) is characterized by one broad peak. It seems that the broad peak above $T_c$ corresponds to the non-resonant component observed below $T_c$ [10].

Fig. 2 summarizes the evolution of the normalized $\chi''(Q_{AF},\omega)$ at 5 K in YBCO for oxygen contents $x$ = 0.5; 0.52; 0.83; 0.92, and 0.97 [5,2]. The typical energy resolution in INS measurements is about of 4 - 5 meV [8,9,11-15]. In Fig. 2, one can see that, below $T_c$, the resonant peak is dominant in the odd channel of the spin susceptibility. The non-resonant component as a shoulder follows the resonant peak with increasing doping. The resonance peak at (π,π) corresponds to the resonant component in the odd channel of the local spin susceptibility [2,3], which we will discuss in the next Section. The energies of these two peaks practically coincide [3].

In the literature, there are many possible explanations for the origin of the commensurate sharp resonance peak (see Introduction). The non-resonance peak in spin-susceptibility spectra at (π,π), shown in Fig. 2, corresponds most likely to incommensurate peaks [10] observed recently below the resonance peak in YBCO$_{6.6}$ [6,15,20]. It is clear that the resonance peak relates to the establishment of SC [2,3,10,12], while it seems that the non-resonant peak does not. The latter corresponds most likely to the AF phase which co-exists with the SC in cuprates [26]. We analyze the spin-susceptibility spectra at $Q_{AF}$ in YBCO and Bi2212 and the origin of the resonance peak in Section 5.

## 4. Q - integrated magnetic spectra

In this Section, we analyze the spin susceptibility integrated over 2D in-plane wave vector, which is also known as local susceptibility, $\chi''_{2D}(Q_z,\omega) \equiv \int \chi''(Q,\omega)\, d^2Q / \int d^2Q$, where the **Q** integrals are over $(Q_x, Q_y)$ only and $Q_z$ is chosen to be close to an optical or acoustic position [2,3,8]. Because the $CuO_2$ planes in YBCO actually appear in coupled bilayers, magnetic fluctuations that are in-phase (acoustic) or out-of-phase (optical) with respect to the neighboring plane will have different spectra. In single-layer LSCO, there is no such splitting in the local susceptibility. Alternatively, the data can be summarized in terms of the peak intensity at $Q_{AF} = (\pi,\pi)$, which was discussed in the previous Section. The local and peak susceptibilities are not proportional to each other, because of the energy dependence of the momentum line shape [2,3]. The $Q$-integration has the effect to change the shape of the spin susceptibility as it enhances the high energy part due to the broadening in $Q$-space of the AF fluctuations [2,3]. However, the resonant component in the odd channel of the local spin susceptibility corresponds to the resonance peak at $(\pi,\pi)$ [2,3].

We turn to the interpretation of INS spectra obtained on LSCO and YBCO with different hole concentrations. Figs. 3(a) and 3(b) depict the local acoustic (odd) and optical (even) susceptibility, respectively, obtained at 5 K on underdoped YBCO with $T_c$ = 52 K [8]. In Figs. 3(a) and 3(b), one can clearly discern two different magnetic contributions. The resonant component appears in the odd channel at $E_r$ = 24 meV while the spin gap occurs in the even channel at $E_{s-g}$ = 53 meV [8]. The broad peak at about 60 meV in the odd channel corresponds most likely to the spin gap from the even channel [8]. The phase diagram shown in Fig. 1 can help to interpret the spin-susceptibility data shown in Figs. 3(a) and 3(b) if we assume that $E_r = 2\Delta_c$ and $E_{s-g} = \Delta_p$. The hole concentration can be approximately calculated from the empirical relation $T_c/T_{c,max} = 1 - 82.6(p - 0.16)^2$ which is satisfied for a number of cuprates [50] and we use $T_{c,max}$ = 93 K for YBCO and $T_{c,max}$ = 38 K for LSCO. For YBCO with $T_c$ = 52 K, it gives $p/p_m$ = 0.54. From the phase diagram shown in Fig. 1, the magnitudes of the two OPs in YBCO at $p/p_m$ = 0.54 are equal to $\Delta_c$ = 12.4 meV and $\Delta_p$ = 54 meV. Thus, there is a good agreement between the two sets of data, namely, that $E_r = 2\Delta_c$ and $E_{s-g} = \Delta_p$. In *undoped* $YBCO_{6+x}$ with $x$ = 0.15 [13] and $x$ = 0.2 [14], the resonant component in the odd channel is absent, and the spin gap is detected in the even channel at 74 and 67 meV, respectively. If one will continue the straight line in the phase diagram shown in Fig. 1 at low hole concentrations, which corresponds to the pairing OP, and assume that $\Delta_p$ = 74 and 67 meV, then, one obtains that $p/p_m$ = 0.04 and 0.16, respectively. These two points $p/p_m$ = 0.04 and 0.16 are, indeed, outside the SC zone in the phase diagram shown in Fig. 1.

From the INS data discussed above, it is clear that the resonant component relates to the establishment of SC since it is present exclusively in the SC state. One may conclude that the spin gap which exists also in *undoped* YBCO has no relations with the SC. However, it may be not the case: the spin gap can be a consequence of the local pairing without long-range phase coherence. Secondly, the result that the resonant component appears in the INS spectrum at the energy equal to the *double* magnitude of coherent OP, $E_r = 2\Delta_c$, which is not the case for the spin gap ($E_{s-g} = \Delta_p$), is also very important. It is very suggestive that the coherent OP has the magnetic origin due to spin fluctuations since the condensation energy of a Cooper pair is, in general, equal to $2\Delta$. We discuss this important result in Section 5.

To our knowledge, there are not so much local-susceptibility data in YBCO presented in the literature. In other two local-susceptibility measurements on YBCO$_{6+x}$ with $x$ = 0.6 ($T_c$ = 62.7 K) [7] and $x$ = 0.7 ($T_c$ = 67 K) [3], the comparison of $E_r$ and $E_{s-g}$ values with the $2\Delta_c$ and $\Delta_p$ magnitudes gives the satisfactory results (the data are presented in Fig. 4).

Let us analyze INS spectra obtained on another hole-doped cuprate, LSCO. Fig. 3(c) shows local-susceptibility data obtained at 17 K on slightly underdoped LSCO ($x$ = 0.14) with $T_c$ = 35 K [9]. The local susceptibility in Fig. 3(c) has only one very broad peak at about 20 meV. At first sight, it seems that the local-susceptibility data in LSCO do not show the presence of the spin gap. However, this fact can be explained by using again the phase diagram. From Fig. 1, for LSCO ($x$ = 0.14) with $T_c$ = 35 K which corresponds approximately to $p/p_m$ = 0.8, one obtains that $2\Delta_c$ = 16.4 meV and $\Delta_p$ = 18.3 meV. Consequently, it seems that the two components (the resonant component and spin gap) are simply superimposed in the INS spectrum shown in Fig. 3(c). This is the reason why the spin gap is "absent" in Fig. 3(c). The hole concentration is the main controlling factor for the shape of spin-susceptibility data. Unfortunately, we are not aware of other published data on the local susceptibility in LSCO.

Very recently, Fong *et al.* considered the temperature evolution of even and odd spectra in undoped YBCO ($x$ ~ 0.5 - 0.7) [3]. The magnetic susceptibility at high temperatures is rather featureless. The spin susceptibility gradually enhances in both even and odd channels at approximately the same rate, but centered around different frequencies. Both spectra just above $T_c$ show a broad peak in each channel that is merely shifted in frequency. At $T_c$, the peak in the odd channel sharpens abruptly in contrast to the even channel which shows no sudden changes. Below $T_c$, the intensity in the even channel continues its smooth normal-state evolution and eventually saturates at low temperature. Thus, the temperature dependencies of the two components in INS spectra point out once again that the resonant component intimately relates to the establishment of the SC which almost does not affect the spin gap.

To visualize, we present the local-susceptibility data in YBCO and LSCO at different hole concentrations immediately in the phase diagram. In Fig. 4, one can see that the resonant-component position, $E_r$, in YBCO (open squares) scales with the magnitude of the coherent OP approximately as $E_r = 2\Delta_c$. The spin-gap data in YBCO (filled squares) in Fig. 4 are in a good agreement with the pairing OP, $E_{s\text{-}g} = \Delta_p$ [51]. In the LSCO ($x = 0.14$), we assume that the two components (the resonant component and spin gap) are superimposed in the local-susceptibility spectrum shown in Fig. 3(c). We use the position of the peak at about 20 meV for the two components simultaneously. The data in the LSCO are shown in Fig. 4 as diamonds.

To summarize, the analysis based on the local-susceptibility data in YBCO and LSCO, published so far, shows that the INS spectra can be interpreted by using the phase diagram of hole-doped cuprates.

## 5. Discussion

We summarize in this Section our interpretation of the INS spectra and suggest an experiment to verify the interpretation.

The resonance peak at $(\pi,\pi)$ relates directly to the resonant component in the odd channel of the local spin susceptibility. Since there are not so much local-susceptibility data available in the literature, in Fig. 4, in addition to the resonant-component position data in the odd channel of the local susceptibility, we present also the resonance-peak energy, $E_r$, in YBCO and Bi2212, discussed in Section 3. In Fig. 4, we use the $E_r$ data from Fig. 2 and other data available in the literature [3,5,7,12,15,17-19,21]. One can see, that, in Fig. 4, there is a good agreement between the resonance-peak position data and double magnitude of the coherent OP, $E_r = 2\Delta_c$. In the phase diagram, we present also the energy of the non-resonant component at $Q_{AF}$ [52]. The data of the non-resonant component in Fig. 4 are taken from Fig. 2 and the literature [3,12]. In Fig. 4, the magnitude of the non-resonant component accounts, on the average, for 70 % of the magnitude of the coherent OP.

It is important to note that the hole concentration, $p/p_\mathrm{m}$, for all INS data shown in Fig. 4 has been calculated using the $T_\mathrm{c}$ value. Consequently, any INS point (dots, squares, triangle, diamonds) in Fig. 4 has a relatively large horizontal error which can be the order of ± 0.04. The vertical errors for the INS data shown in Fig. 4 are defined by the resolution in INS measurements, which is the order of 4 - 5 meV [8,9,11-15]. However, in spite of the errors, the INS data shown in Fig. 4, in general, reflect the tendencies which are used in the present work for the interpretation of the magnetic spectra.

We turn now to the interpretation of the spin-susceptibility data. The non-resonant peak in spin-susceptibility spectra at (π,π), shown in Fig. 2, corresponds most likely, to incommensurate peaks [10] and, probably, does not relate directly to the establishment of the SC. From Fig. 4, it is clear that the resonance peak (the resonant component in the local susceptibility) relates to the coherent OP. In the literature, there are many possible explanations for the origin of the magnetic resonance peak (see Introduction). The most popular explanation is that the magnetic resonance peak is due to electron-hole pair excitation across the SC energy gap [3,10]. However, it is implausible. Any SC energy gap has the temperature dependence which is *somehow* similar to the BCS temperature dependence [40]. According to any explanation for the presence of the resonance peak in INS spectra, which involves the SC gap, by increasing the temperature the resonance-peak position has to shift from the maximum energy to zero. However, it is not the case. The position of the resonance peak does not depend on temperature [3,7]. Even, if we consider the case in which the SC energy gap is closing above $T_\mathrm{c}$, INS measurements show that the position of the resonance peak above $T_\mathrm{c}$ has a weak dependence on temperature [3,7]. Thus, any explanation which involves directly the SC energy gap is implausible. However, the temperature dependence of the intensity of the resonance peak is very similar to the temperature dependence of a SC gap [7]. By analyzing tunneling and INS data [36] we proposed that the resonance peak corresponds to spin-waves into local AF domains, which mediate the long-range phase coherence of the spinon SC which occurs along charge stripes (a Magnetic Coupling between Stripes (MCS) model) [37-39]. Spin-waves are excited by dynamically fluctuating charge stripes. The SC mediated by spin fluctuations implies that its OP has the $d_{x^2-y^2}$ (hereafter, d-wave) symmetry [32,53,54]. Indeed, there is a consensus that the predominant OP in hole-doped cuprates has the d-wave symmetry [55]. In our previous studies, we considered in detail the symmetries of the two OPs shown in Fig. 1. We found that the OP for long-range phase coherence in hole-doped cuprates, indeed, has most likely the d-wave symmetry [46,38]. By other words, we assume that the resonant-peak position corresponds to the carrier frequency of magnons which mediate the electron pairing in the d-wave channel. The carrier frequency of magnons does not depend or depends slightly on temperature. The main factors which are responsible for the temperature dependence of the magnitude of the coherent OP are the intensity of the resonance peak and a hole-spin-waves coupling constant [54]. It is possible that the latter depends also on temperature.

Some theoretical calculations show that spin fluctuations are too weak to produce the pairing mechanism in the cuprates. However, the comparison of INS and infrared spectroscopy data shows that (i) the conducting carriers are strongly coupled to a resonance peak, and (ii) the coupling strength inferred from those results is sufficient to account for high value of $T_c$ [58].

Let us now summarize our interpretation of the INS data in LSCO, YBCO and Bi2212. In the cuprates, the SC condensate is not able to establish the coherent state by itself due to a short coherent length and its low density. We assume that it uses spin fluctuations into local AF domains to establish the long-phase coherence. Thus, in frameworks of our interpretation, this is the reason why, in INS measurements, local-susceptibility data show a *spin gap* with the magnitude equal to the magnitude of the pairing OP, $\Delta_p$, and *magnon peak* at the energy equal to the double magnitude of the coherent OP, $2\Delta_c$. Since an INS measurement is the signal averaging experiment two (or more) magnetic signals coming from different domains of a sample are superimposed, and since the magnitudes of the two OPs in hole-doped cuprates have different dependencies on hole concentration (see Fig. 1) INS spectra look differently depending on *p*. For the local susceptibility, there are three different cases: (i) In the strongly underdoped regime where $\Delta_p > 2\Delta_c$, the magnon peak occurs inside the spin gap. (ii) In slightly underdoped regime where $\Delta_p \approx 2\Delta_c$, the two signals appear at the same energy and, in the case of LSCO, can not be distinguished. (iii) At optimal doping and in the overdoped regime where $\Delta_p < 2\Delta_c$, the magnon peak appears above the spin gap. Magnons which propagate along the AF wave-vector mediate the d-wave pairing. From the analysis made in our previous studies, the pairing OP has most likely an anisotropic s-wave symmetry [46,38]. This can explain the shape of spin-gap data in INS spectra.

It is widely believed that the $Nd_{2-x}Ce_xCuO_4$ (NCCO) cuprate (and its homologues) is electron-doped since it shows a clear *n*-type Hall coefficient through most of the doping level [59]. There is a consensus that the total OP in NCCO has the s-wave symmetry [60]. This suggests that the SC mediated by spin fluctuations is absent in NCCO [54]. Pulsed neutron scattering measurements found no resonance peak in NCCO [61]. The latter fact also demonstrates that the magnetically mediated SC is absent in NCCO. The absence of SC mediated by spin fluctuations in NCCO is, in fact, surprising since spin waves do propagate into $CuO_2$ planes in undoped NCCO ($x \leq 0.14$) like in other cuprates [25-27]. However, the increase of Ce concentration above $x = 0.14$ softens strongly spin waves in NCCO [27]. The s-wave symmetry of total OP in NCCO implies that both the pairing OP and OP for long-range phase coherence have a s-wave symmetry.

To verify our interpretation of the INS data we propose to perform local-susceptibility measurements in optimally doped YBCO. According to our scenario, the INS spectra will show that the magnitude of the spin gap is less than the position of the resonance component.

Finally, in spite of the fact that it is not the purpose of the present work, it is important to notice that, in frameworks of our interpretation of the INS data (not the resonance peak), we find that the pairing OP is a pure spin gap. The latter means that the Cooper pairs in cuprates consist of neutral fermions, *i.e.* spinons. In fact, through recent INS measurements, Lake *et al*. made the same conclusion [22].

## 6. Conclusions

In summary, we discussed the $Q$-integrated INS data and data obtained at AF vector, $Q_{AF} = (\pi,\pi)$, in LSCO, YBCO and Bi2212. We showed that by using the phase diagram of hole-doped cuprates it is possible to interpret the INS data: the spin gap which appears in the even channel of the local susceptibility corresponds to the pairing OP while the resonance component in the odd channel relates directly to the coherent OP. We discussed also the origin of the magnetic resonance peak which, most likely, corresponds to spin-waves propagating into local AF domains and mediating the long-range phase coherence in hole-doped cuprates.

## Acknowledgements

I thank K. Yamada, M. Arai and R. Deltour for discussions. This work is supported by PAI 4/10.

FIGURE CAPTIONS:

Fig. 1. Phase diagram of hole-doped cuprates: $\Delta_c$ is the magnitude of the coherent OP, and $\Delta_p$ is the magnitude of the pairing OP. The $p_m$ is a hole concentration with the maximum $T_c$ (from Ref. 45).

Fig. 2. Imaginary part of the odd spin susceptibility at 5 K in YBCO versus $x$ in the SC state (from Refs. 5 and 2). Note, the spectra are average, measured points are not shown.

Fig. 3. Average odd (a) and even (b) spin susceptibilities in YBCO at 5 K (from Ref. 8). (c) Average local susceptibility in LSCO at 17 K (from Ref. 9). In all frames, measured points are not shown.

Fig. 4. Phase diagram of hole-doped cuprates (see Fig 1) and INS data at low temperature: squares and dots (YBCO), diamonds (LSCO) and triangle (Bi2212). The filled diamond and filled squares are spin-gap data in the even channel of the local susceptibility, multiplied by 2. The open diamond and open squares are resonant-component energy data in the odd channel of the local susceptibility. The triangle and filled dots show the energy of the magnetic resonance peak in YBCO [2,3,5,7,12,15,17-19] and Bi2212 [21] at different hole concentrations. The open dots show the energy of non-resonant component in INS spectra at (π,π) in YBCO [2,3,5,12] (the data are not multiplied by 2 [52]). The diamonds correspond to a peak of the local susceptibility in LSCO [9], assuming that two signals are superimposed, (see text). The data in YBCO and Bi2212 are measured at 5 K, the data in LSCO are obtained at 17 K. For the INS data, the hole concentration was calculated using the $T_c$ value (see text). The dash line is a guide to the eye.

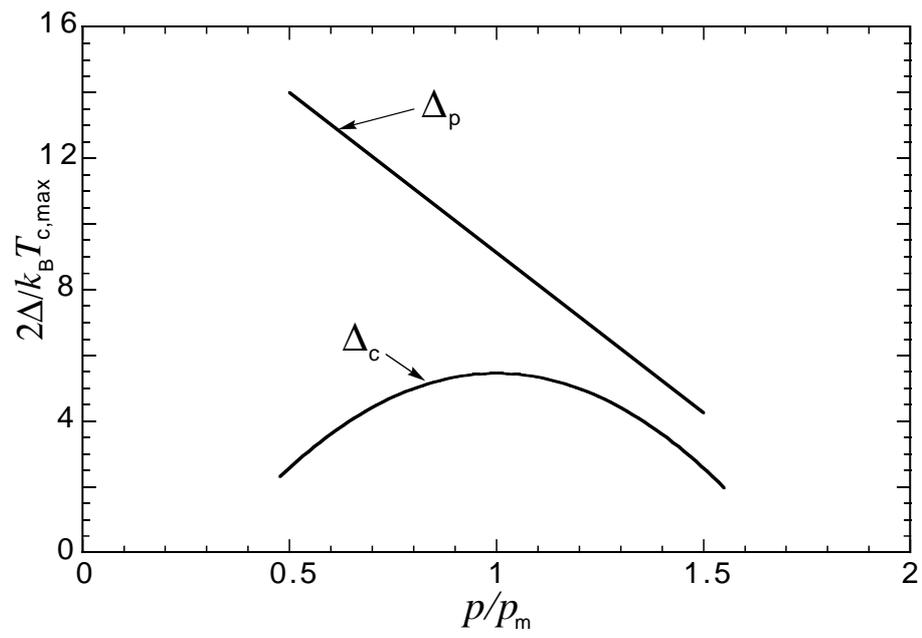

FIG. 1

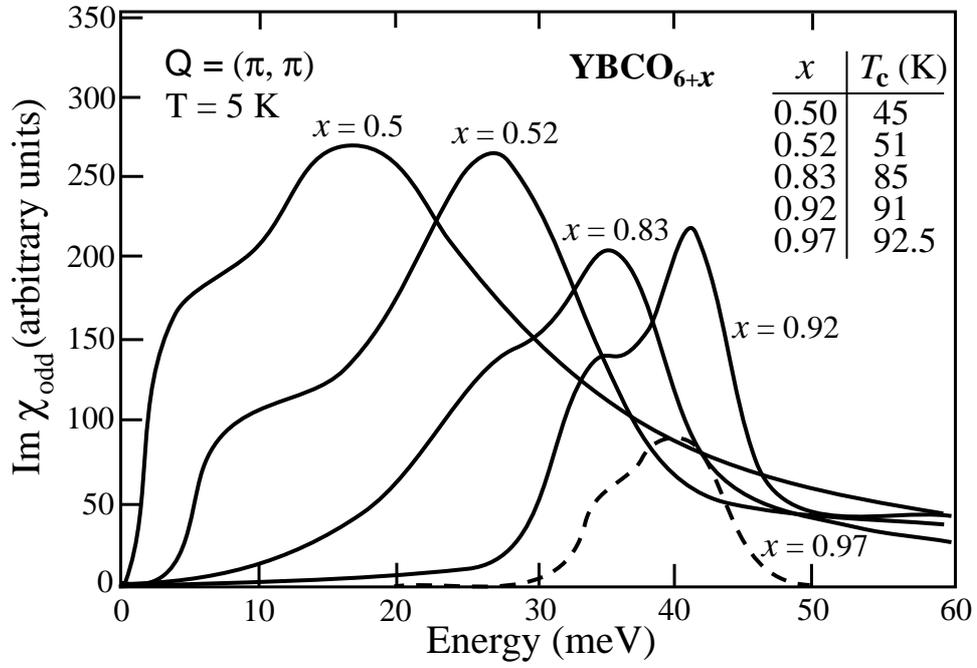

FIG. 2

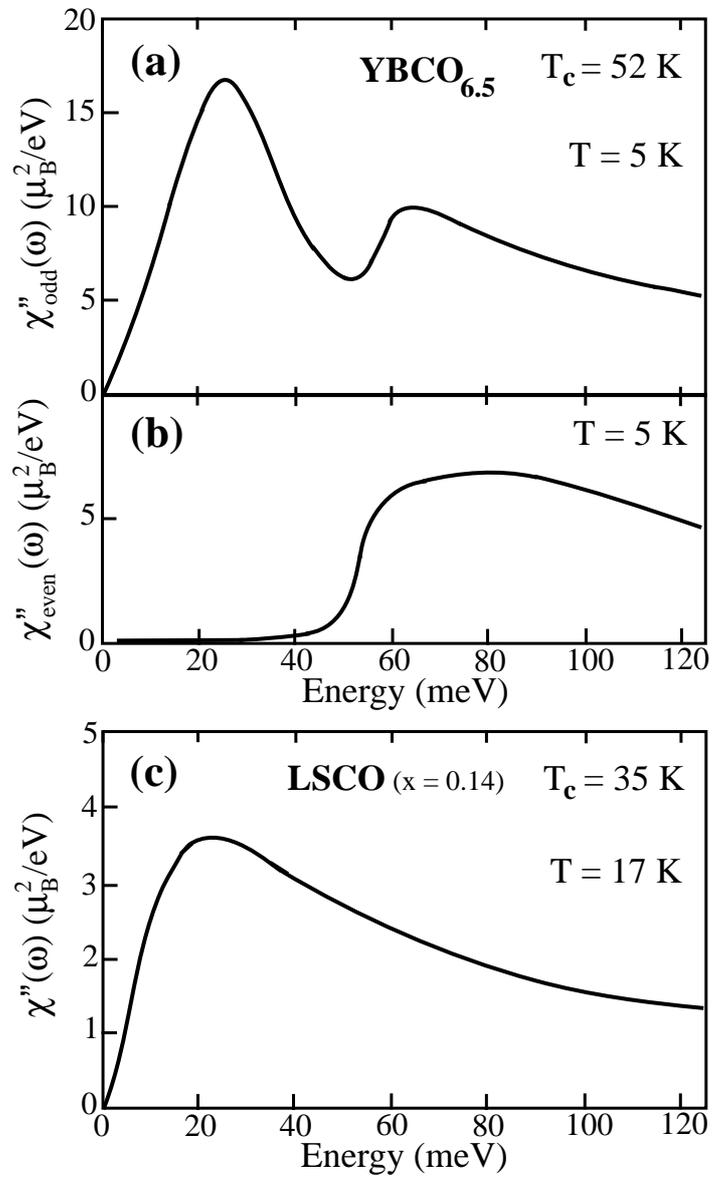

FIG. 3

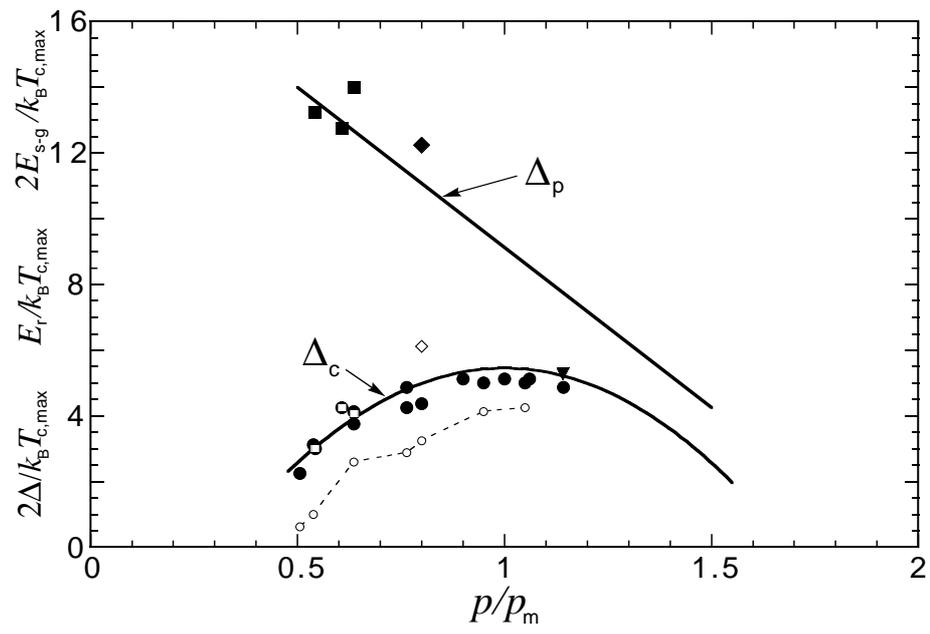

FIG.4